\documentclass{llncs}

\usepackage[english]{babel}
\usepackage{blindtext}

\usepackage{lipsum}
\usepackage{booktabs}
\usepackage{paralist}
\PassOptionsToPackage{hyphens}{url}
\usepackage{hyperref}
\hypersetup{colorlinks, citecolor=blue, filecolor=blue, linkcolor=blue, urlcolor=blue}
\usepackage{multirow}
\usepackage{subcaption}
\usepackage{graphicx}
\usepackage{xspace}
\usepackage{xcolor}

\usepackage[inline]{enumitem}

\usepackage{tikz}

\newcommand\encircle[1]{%
	\tikz[
		baseline={([yshift=-8pt]current bounding box.north)}
	]
		\node (X) [draw, shape=circle, inner sep=0, fill=black, text=white,scale=0.7] {\strut #1};
	\hspace{-4pt}
}

\newcommand{\afblock}[1]{\noindent{\textbf{#1}}}
\newcommand{\takeaway}[1]{\noindent{\textbf{Findings.}} \textit{#1}}

\usepackage{todonotes}

\newcommand{\ie}{i.e.,\,}
\newcommand{\Ie}{I.e.,\,}
\newcommand{\eg}{e.g.,\,}
\newcommand{\Eg}{E.g.,\,}
\newcommand{\etal}{et~al\@ifnextchar.{}{.\@}}
\newcommand{\etc}{etc\@ifnextchar.{}{.\@}}

\newcommand{\wrt}{w.r.t.\@}
\newcommand{\sref}[1]{\S~\ref{#1}}

\begin{document}

\def\UrlBreaks{\do\/\do-}

\title{Differences in Social Media Usage Exist Between Western and Middle-East Countries
}

\author{%
Jens Helge Reelfs\inst{1} \and 
Oliver Hohlfeld\inst{1} \and
Niklas Henckell \inst{2} 
}%
\institute{
Brandenburg University of Technology, Cottbus, DE\\
\email{\{lastname\}@b-tu.de}\\ 
\and
The Jodel Venture GmbH, Berlin, DE\\
\email{niklas@jodel.com}
}

\maketitle

\vspace*{-2em}
\begin{abstract}
In this paper, we empirically analyze two examples of a Western (DE) versus Middle-East (SA) Online Social Messaging App.
By focusing on the system interactions over time in comparison, we identify inherent differences in user engagement.
We take a deep dive and shed light onto differences in user attention shifts and showcase their structural implications to the user experience.
Our main findings show that in comparison to the German counterparts, the Saudi communities prefer creating content in longer conversations, while voting more conservative.
\end{abstract}
\section{Introduction}

Every social networking platform depends on an active user-base.
The recent pandemic disruption has even accelerated the shift of our everyday's life into digital spaces, may it be at work or socially.
What constitutes an active user-base and if cultural differences exist in this usage behavior, however, is unclear.

Social network analysis is an active field of research for more than a decade.
Research provided a general understanding through the empirical and qualitative analyses of a number of different networks.
Examples include structural measurements of classic online social networks \cite{mislove2007measurement,nazir2008unveiling,schioberg2012tracing,kairam2012talking} as well as more specialized variants such as microblogging~\cite{bollen2011modeling}, picture sharing~\cite{vaterlaus2016snapchat,cha2009measurement}, or knowledge sharing~\cite{wang2013empirical}.
Works in this field analyzed the networks' {\em structure}, mostly using graph-theory approaches. %
This way, they showed that social networks usually creates small-world networks~\cite{manku2004know,freeman2004development}.
The influence of cultural or geographic backgrounds on usage largely remain unknown.

Most studies either focus on analyzing social-media usage \emph{worldwide} or by focusing on specific parts of the world, mostly English speaking.
These works have enriched our understanding of social media.
Yet it is unclear if or to what extent the obtained measures differ between different geographic regions.
Cultural differences are known to exist that drive human behavior in social networks, e.g., the degree of connectivity~\cite{Overload} or how marketers use social media to impact purchase decisions~\cite{HowSocial}.
Yet, little is known on how geographic or cultural backgrounds may impact the way users interact with a social media platform in terms of the generated traffic; that is, content creation and content voting.

In this work, we take the rare chance to analyze ground truth information provided by a social network operator to compare interactions with a social media platform of a Western (Germany) and a Middle-Eastern (Saudi Arabia) country.
We selected both countries since they represent the largest user-bases of the Jodel platform, while simultaneously representing a largely different (cultural) background.
Our data sets capture the entirety of all social media interactions in both countries since the very first post.
This way, we can, for the first time, shed light on whether geographic or culture specific differences exist between both countries \wrt{} how the user-bases generate and vote content.

The studied social media platform Jodel is location-based and anonymous.
Most importantly, the feature of Jodel to form independent local communities enables us to compare in-country and between country effects and thereby to clearly identify country specific usage differences.
Further, it does not display any form of user profiles or other user-related information that would introduce visible social credit; users solely interact framed into their physical proximity and based on their topic preferences.
This results in a \emph{pure} form of communication that is reduced to content, since any form of influence by user profiles such as social status is absent.
This makes Jodel an ideal platform to study differences in content creation and voting, \ie{} the entirety of active interactions with this social network.
We shed light on fundamentally different user behavior and engagement patterns within such anonymous spaces having received less attention as of today, across the Kingdom of Saudi Arabia and Germany; and showcase structural implications.
Our contributions are as follows.
\begin{itemize}
    \item While not a primary focus of our work, we empirically show the very different adoption processes of a new social media platform in both countries.
    \item We show that, invariant to time and community size, users in Saudi-Arabia (Middle-East) behave fundamentally different to the German counterparts (Western country). They prefer creating content, but vote slightly less than the German users. This highlights, for the first time, that country-level differences in the usage culture in social media exist that create drastic differences in user behavior of the very same social media platform. 
    \item We exemplify the implications of shown user engagement to the social media platform. \Eg{} we show that the availability of more content in Saudi Arabia naturally decreases the available votes per post, which can have serious impact on, \eg{} distributed community moderation techniques.
\end{itemize}

\section{The Jodel App}
\label{sec:jodel}
\begin{figure}[htb]
    \vspace*{-1em}
	\centering
	\includegraphics[width=0.65\linewidth]{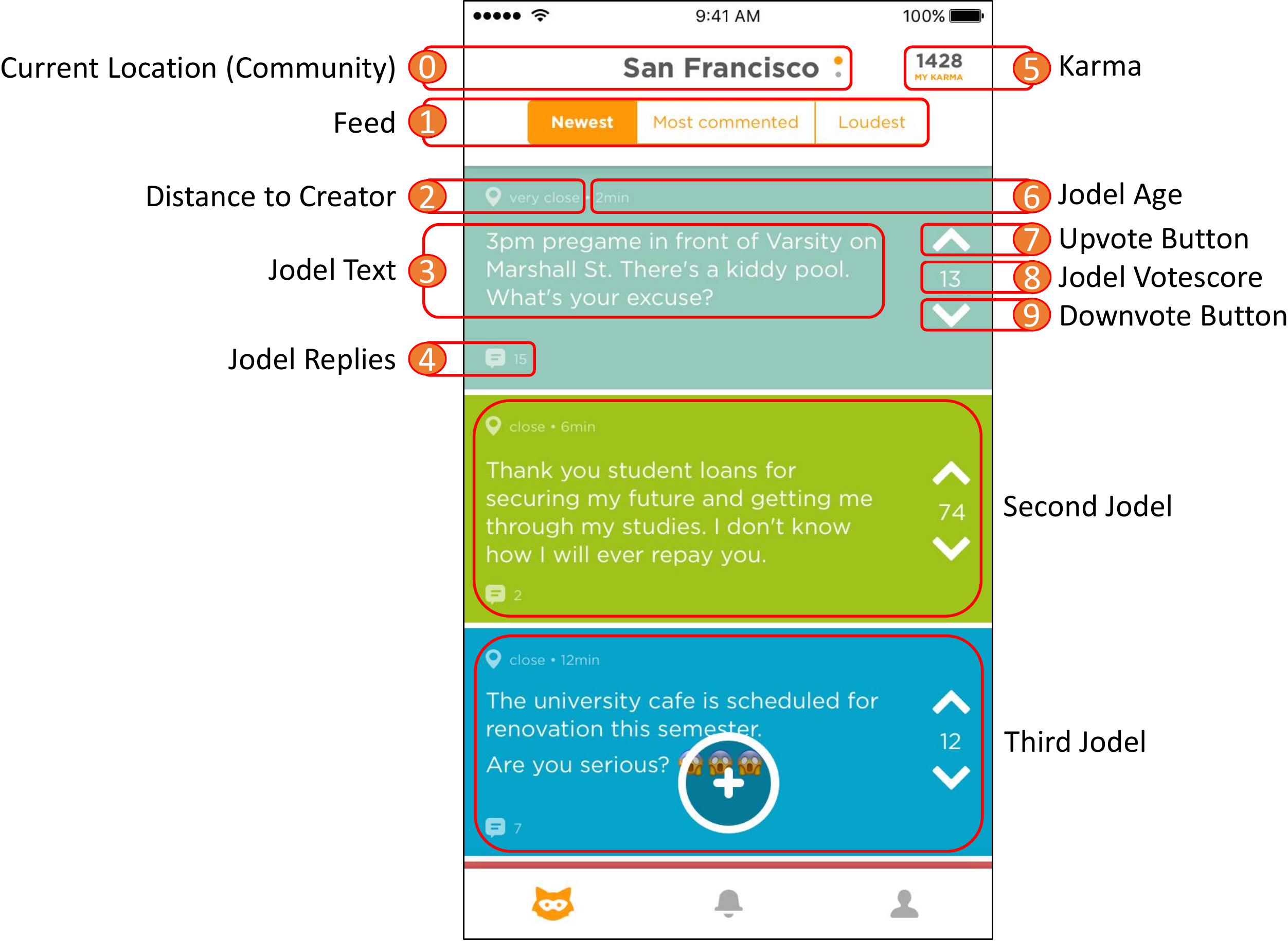}
	\caption{
		\textbf{Jodel iOS mobile application.}
	}
	\label{fig:jodelapp}
    \vspace*{-2em}
\end{figure}

Jodel
is a mobile-only messaging application which we show in Fig.~\ref{fig:jodelapp}.
It is location-based and establishes local communities relative to the users' location \protect\encircle{0}.
Within these communities, users can {\em anonymously} post both images and textual content of up to 250 characters length \protect\encircle{3} (\ie{} microblogging) and reply to posts forming discussion threads \protect\encircle{4}.
Posted content is referred to as ``Jodels'', colored randomly \protect\encircle{3}.
Posts are only displayed to other users within close (up to $\approx$20km) geographic proximity \protect\encircle{2}.
Further, all communication is {\em anonymous} to other users since no user handles or other user-related information are displayed.
Only {\em within} a single discussion thread, users are enumerated to enable referencing to other users.
Threads are displayed to the users in three different feeds \protect\encircle{1}: i) \emph{recent} showing the most recent threads, ii) \emph{most discussed} showing the most discussed threads, and iii) \emph{loudest} showing threads with the highest voting score.

Jodel employs a community-driven filtering and moderation scheme to avoid adverse content.
For an anonymous messaging app, community moderation is a key success parameter to prevent harmful or abusive content.
In Jodel, content filtering relies on a distributed voting scheme in which every user can increase or decrease a post's vote score by up- (+1) \protect\encircle{7} or downvoting (-1) \protect\encircle{9} (similar to StackOverflow).
Posts reaching a cumulative vote score \protect\encircle{8}\,below a negative threshold \mbox{(\eg{} -5)} are no longer displayed.
Depending on the number of vote-contributions, this scheme filters out adverse content while also potentially preferring mainstream content.
To increase user engagement \wrt{} posting and voting, Jodesl uses lightweight gamification by awarding \emph{Karma} points \protect\encircle{5}. %
\section{Dataset Description and Ethics}
	\label{sec:Dataset_Description_and_Statistics}
	The Jodel network operators provided us with {\em anonymized} ground truth data of their network, cf. Table~\ref{tab:dataset}.
	The obtained data contains post and interaction \emph{metadata} created within Germany and the KSA only, and spans over multiple years from the beginning of the network in 2014 up to August 2017.
	It is limited to metadata only without textual content and user records stripped and anonymized.
	The data enables us to cluster users by their anonymous ID.
	Further, it contains no personal information and cannot be used to personally identify users.
	While users consent to research via the Jodel ToS, we also inform and synchronize with the Jodel operator about evaluations we perform on their data.
	The structure of our available dataset includes 3 categories: interactions, content, and users.

	\begin{table}[t]
		\vspace*{-1em}
		\small
		\centering
		\begin{tabular}{l|r|r|l}
			\toprule
			\textbf{Type}		& \textbf{\#SA}	& \textbf{\#DE}		& \textbf{Description}\\ \midrule
			User				& 1.2M	& 3.6M	& User metadata\\
			Content				& 469M	& 285M	& Content (posts, replies)\\
			Interaction			& 961M	& 3.0G	& Interactions incl. user, community\\
								&		&		& and type (post, reply, up-/downvoted)\\
			\bottomrule
		\end{tabular}
		\caption{
			\textbf{Dataset statistics.}
			The data ranges from the application start in late 2014 up to the beginning of August 2017. We find the first observation in the KSA in December 2014.
		}
		\label{tab:dataset}
        \vspace*{-3em}
	\end{table}

	\afblock{Dataset limitations.}
	Our dataset only includes the users' {\em active} interactions with the system, where they contribute like registering, creating posts, replying, or voting.
	Thus, we cannot infer when or how much a user only {\em passively} participates---lurkers---who only consume content.
	Further, the vote interactions are always mapped to the date and community of the respective content creation.
	This prevents us from making detailed analyses depending on the voting time or place.
	However, due to the vivid usage of the application (multiple posts/replies per minute), we generally consider votes to be executed on the same day as their respective content.
	Especially since posts are only accessible via the three different feeds, where they will only stay for a very limited time, casting votes long after the content creation is usually not possible.
\section{The Birth of the Jodel Networks in DE and the KSA}
\label{sec:growth}

The growth patterns of social networks are less understood given that data captures from the very beginning of a social media platform are typically unavailable.
We take the rare chance to begin showcasing the rate in which the Jodel platform established itself in both countries.
Our first peek is relevant to better understand network activity and to define a meaningful aggregate layer for comparison, \eg{} time slices, for studying cultural shifts in social media usage in the next section.
We thus remark that a detailed study of the shown adoption pattern is beyond the scope of this paper.

\subsection{Different Adoption Pattern in Germany and the KSA}

\begin{figure}[t]
    \vspace*{-1em}
    \centering
    \includegraphics[width=.90\linewidth]{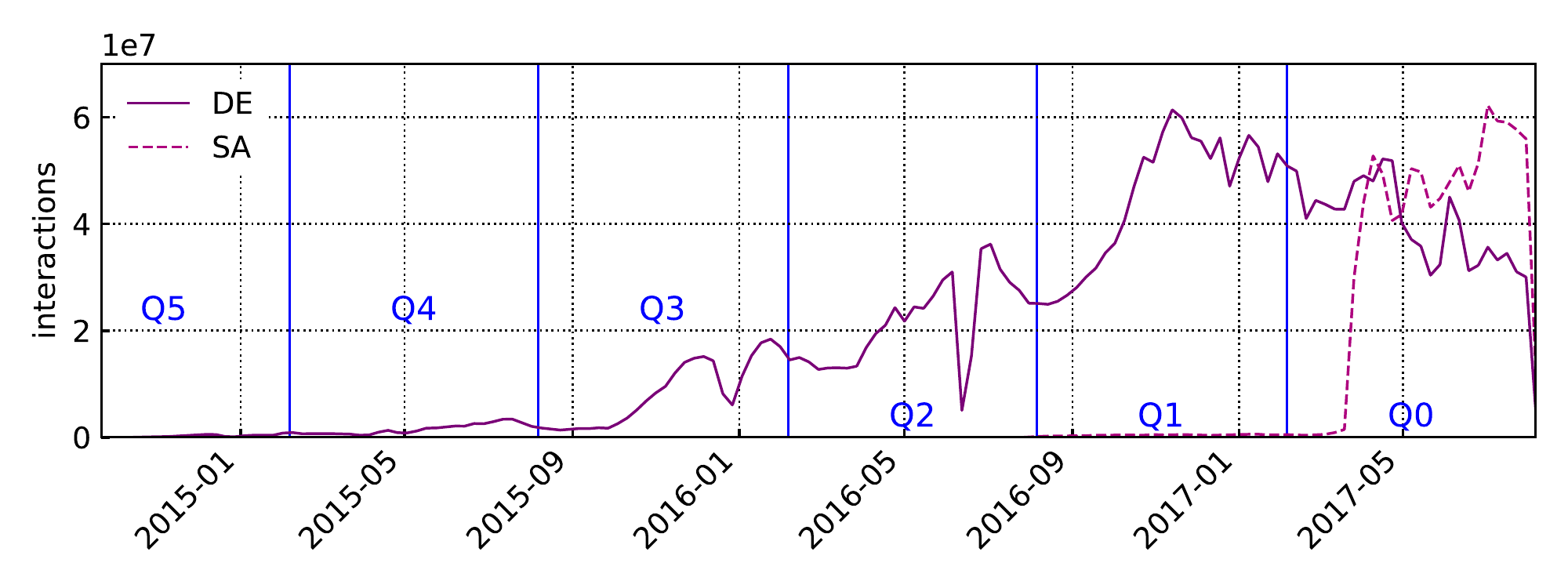}
    \vspace*{-1.5em}
    \caption{Jodel activity in Germany (DE) and the Kingdom of Saudi Arabia (SA) over our observation time. While the DE network enjoys a steady growth over time, the usage in the KSA has suddenly surged and remains stable.}
    \label{fig:timeline}
    \vspace*{-2em}
\end{figure}

We show the adoption of the Jodel network in both networks by the number of interactions over time in Figure~\ref{fig:timeline}.
The figure shows the number of weekly interactions for Germany (solid line) and the KSA (dotted) since the very first interaction till the end of our data set in August 2017.
With interaction, we refer to any interaction with the Jodel system, \ie{} either posting, replying, or voting.

The adoption of Jodel in Germany is characterized by a slow but rather steady growth of network activity over time, peaking in 2016/2017.
This captures the birth of the Jodel network that originated in Germany and then constantly increased in popularity.
In contrast to the steady increasing activity in Germany, the adoption of Jodel in the KSA is characterized by a substantial influx of users and an increase in activity at a short time in March 2017.
To our best knowledge, the reason for this behavior is that Jodel went viral via social media in Saudi Arabia---in absence of any marketing campaigns of the operator itself.

While studying the reasons that were driving these adoption processes is beyond the scope of this paper, their adoption processes differ substantially.
That is, referring to Table \ref{tab:dataset}, in only 4 months, Jodel KSA has roughly gathered 1.2 million users, while Jodel Germany over six times longer time period accumulates to 3.6 million.
Likewise, the amount of interactions equally scales between the KSA with 1 billion and Germany with 3 billion interactions.
This observation and differences in adoption allow for putting aggregates, \eg{} comparable time slices, for our study into perspective.

\takeaway{Adoption pattern and thus associated traffic can differ substantially.}

\subsection{Different Adoption Pattern Require Comparable Time Slices}
\label{sec:slicing}

To compare the social media usage in both countries, we need to derive comparable datasets.
That is, we need to enable comparisons between two different populations of posts, interactions, and users across a multitude of communities.
We thus identify two main dynamics within the lifecycle of communities defining the aggregation dimension:
\emph{i)} time and 
\emph{ii)} per community interaction volume.

\afblock{Slicing by i) Time.}
\label{sec:partition_time}
As we are limited in the length of observations, especially for Saudi Arabia, we selected half-year long timeframes backwards from the end of observation.
We show these timeframes with vertical bars in Figure~\ref{fig:timeline} named by country \emph{DE0, DE1, ..., DE5, SA0, SA1}, where the index indicates the partition's age.
This simple batching approach creates half-year slices that represent various stages within the community development.
In our study, we compare these slices independently in each evaluation to account for the different adoption processes.
We have experimented with higher resolutions to enrich our results with more data points (not shown), but our conclusions remain the same for our presented period length.
We handle these partitions independently of each other, \ie{} early-day users from DE5 may drop out of the statistics in subsequent partitions due to a lack of interactions.

\afblock{Slicing by ii) Community Interaction Volume.}
\label{sec:partition_communities}
Defining ``a'' community is not possible on Jodel given that content is displayed relative to the users location and thus differs from user to user.
That is, every user might experience a slightly different community to interact with, which cannot be reconstructed from the data.
To solve this, we assign each interaction to a nearby major city or district, which generates clusters of interactions that we refer to as communities.
This discretization generates an approximation of the individually experienced communities.
The resulting approximation is of sufficient accuracy to study and compare the Jodel usage in different parts of the respective countries.
Further, the discretization does not normalize for covered area, nor covered population.

We mitigate these inherently hard problems in normalization by simplifying our partitioning approach.
By slicing all interactions into quantiles ordered by their corresponding community size, we enable a relative comparison; named \emph{is\_q0\_25, ...} representing the corresponding quantile of all interactions, discretized into communities (leading to an approximation).
We provide details of this partitioning in Table \ref{tab:city_aggregattion} describing the amount of discretized communities per country.
That is, \eg{} the single largest SA community is the capital Riyadh at about 30\% total interaction volume---hence, it is the only community within the set of \emph{q75\_100}. 
Due to dividing the interaction volume into equal parts, we encounter a heavy-tailed interaction distribution across communities resulting in only few entries within the upper quantiles; the German community size distribution qualitatively matches the SA counterpart, while the latter is largely shifted in volume within magnitudes of fewer communities (not shown).

We will compare the social media usage based on the resulting data sets.

\begin{table}
    \vspace*{-1em}
    \small
    \centering
    \begin{tabular}{c|r|r}
        \toprule
        Interactions quantile  & \#communities SA & \#communities DE\\
        \midrule
        $\approx$ 75..100               & 1   & 14\\
        $\approx$ 50..75                & 4   & 29\\
        $\approx$ 25..50                & 12  & 114\\
        $\approx$ 0..25                 & 78  & 6,678\\
        \bottomrule
    \end{tabular}
    \caption{Interaction volume aggregation layer \& amount of corresponding communities. Due to the heavy tailed distribution across the community discretization, the upper quantiles consist of fewer communities.
    }
    \label{tab:city_aggregattion}
    \vspace*{-4em}
\end{table} %
\section{Geographic Differences in Jodel Usage: DE vs.\ KSA}
\label{sec:cultural_shifts}
Is there a systematic difference between Jodel users in Germany and in the KSA in the way they use the social media platform?
In other words, do culture or country specific usage behaviors exist that uniquely define traffic profiles of the very same platform in each country?
While social media usage has widely been studied, the question of in-platform variation and behavioral differences is still open.
In this section, we set out shedding light on this aspect by comparing the Jodel usage in two countries with a different social and cultural background.

We study the question of (cultural) differences \wrt{} user behavior in Jodel usage by investigating differences in active user \emph{interactions} with Jodel, \ie{} posting and voting, accounting for all possible active system interactions.
We base this evaluation on two factors introduced in Section~\ref{sec:slicing}: partitioning by \emph{i)}~time and \emph{ii)} interaction type, which we discuss next.

\begin{figure}
    \vspace*{-2em}
    \centering
    \begin{subfigure}[t]{.31\textwidth}
        \centering
        \includegraphics[width=\textwidth]{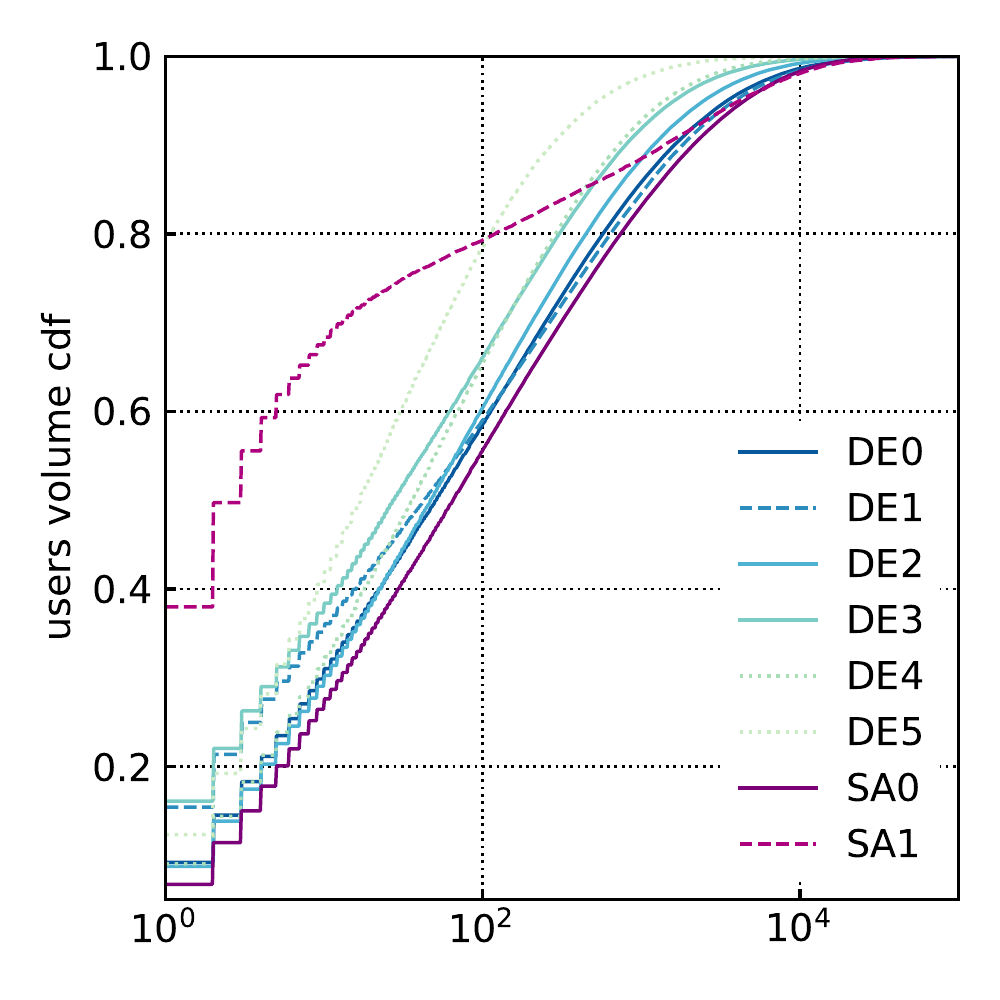}
        \caption{Total per user interactions per time period. Per user interaction volume is very similar over time.}
        \label{fig:user_cdf}
    \end{subfigure}
    \quad
    \begin{subfigure}[t]{.64\textwidth}
        \centering
        \includegraphics[width=\textwidth]{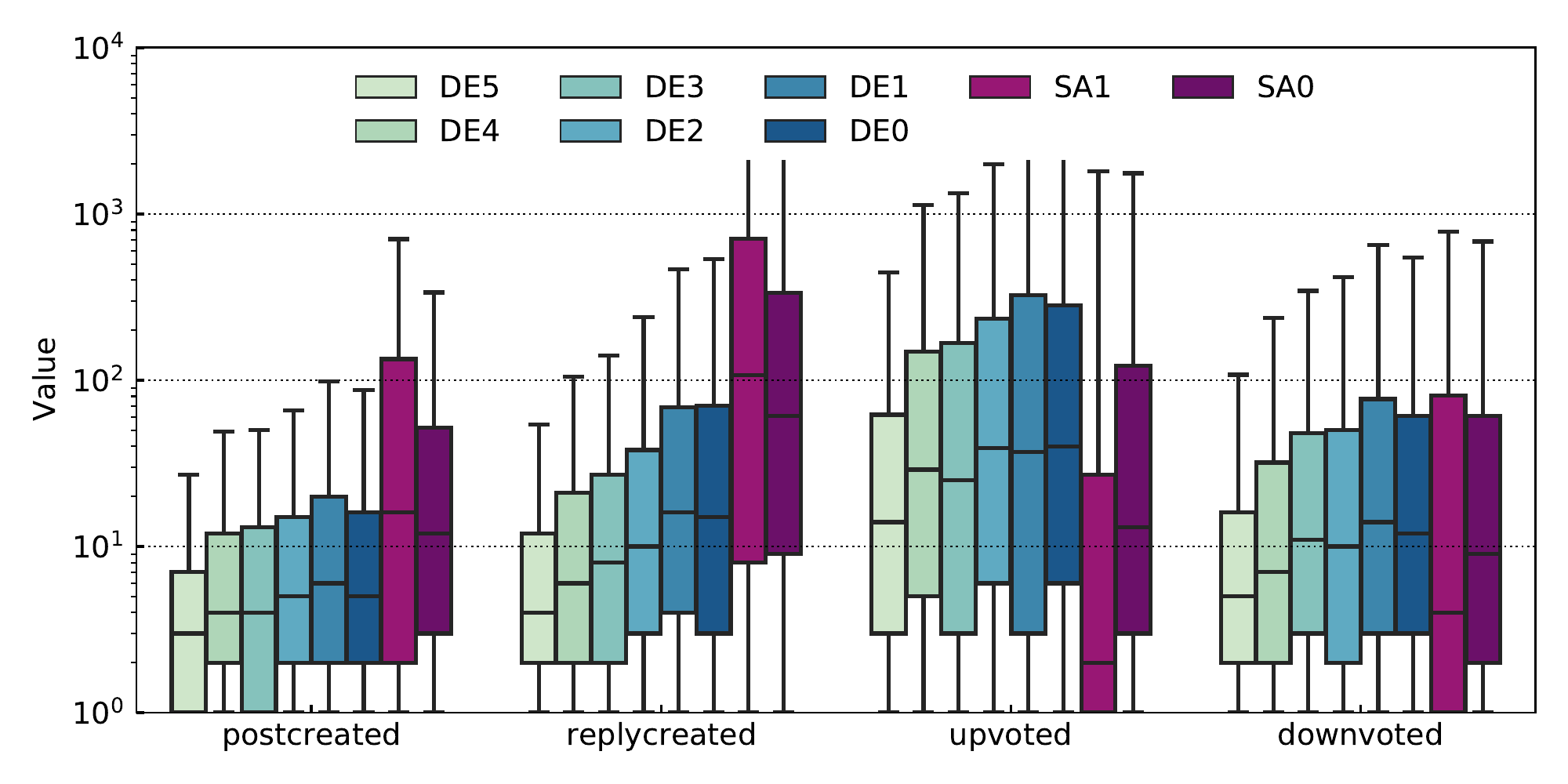}
        \caption{Per user interactions by time period (DE0, ...) and interaction type. Over time, we observe increasing engagement. There is a substantial difference in content creation and upvote activities between SA and DE.}
        \label{fig:user_boxplot}
    \end{subfigure}

    \caption{Differences in per-user interactions by time and country. \emph{a)} While the individual users behave similar across the board \wrt{} interaction volume, \emph{b)} attention shifts towards content creation for the SA user groups, whereas downvoting is less popular in comparison to DE.}
    \label{fig:users}
    \vspace*{-2em}
\end{figure}

\afblock{Overall per-user activity is country-independent.}
First, we subdivide all interactions into independent half-year periods as described in Section~\ref{sec:slicing} (DE0, $\ldots$, DE5, SA0, SA1).
These independent periods enable to compare the behevior of the networks at different times to account for differences in the adoption of Jodel in DE and SA.
We then aggregate the interactions to each user and present the resulting interactions per user CDF in Figure~\ref{fig:user_cdf}.
We show each half-year period as separate CDF (series) for each of the two countries.
Invariant to time, we observe quite similar heavy tailed distributions; that is, most users are not very active, such that \eg{} 60\% of all users each have up to only 100 interactions.
The distribution for SA1 deviates from the pattern since it captures a timeframe before being popular (considerably fewer data points).
In general, we observe that irrespective of time and country, users follow a similar usage behavior---also in absolute terms (not shown).

\afblock{Difference: posting vs. voting.}
Next, we further partition the data by the type of interaction in addition to the time slices used before.
That is, we show distributions of interactions per user subdivided into the voting interactions (upvoted, downvoted) and content creation interactions (postcreated, replycreated) as a box plot in Figure~\ref{fig:user_boxplot}.
Note the logarithmic y-axis.
Further, the whiskers denote the 5\%/95\% percentile.

German users tend to increase their engagement over time at increasing platform activity regardless of the interaction type.
While upvoting is the most prominent type of interaction for the German users, voting content down and replying to content are roughly equally less prominent.
The SA users prefer content creation, especially replying, whereas upvoting happens less frequently.

Remarkably, all time periods within a country are determined by similar behavior.
In other words, posting content is the dominant type of interaction in the KSA, while it is voting in Germany.
This represents a clear difference in platform usage that can be observed between these two countries (also regardless of community size, not shown).

The ratio of up to overall votes remains positive at a happyratio (upvotes to total votes) of 83\% for DE and 71\% for SA.
The figures for the SA1 partition need to be taken with a grain of salt due to only few users; however, the engagement spread is higher compared to the latest timeframe SA0.

\takeaway{Invariant to time and community size, the SA users (Middle-East) behave fundamentally different to the DE counterparts (Western country).
They heavily prefer creating content, but vote slightly less than the German users.
This highlights, for the first time, that cultural patterns in social media user behavior exist that create drastic shifts in how a very same social media platform is used in each country.
This finding may be considered even more interesting, given Jodel being an anonymous platform that enables a very pure form of communication; it entirely focuses on posted content in absence of any user profile.}

\section{Structural Implications}
\label{sec:structural_implications}

With the identified fundamental attention shift in user behavior between SA and DE users in content creation and voting, we now aim at studying resulting implications on the platform.
According to the operator, the communities in both countries are considered to be well-functioning.
That is, regardless of implications arising from different usage profiles, participants are enjoying spending time on the platform to the most part; \eg{} by creating content, voting, or just lurking.

\subsection{Content Voting}
\label{sec:structural_implications__voting}

\begin{figure}[t]
    \vspace{-1.5em}
    \centering
    \begin{subfigure}[t]{.3\textwidth}
        \centering
        \includegraphics[width=\textwidth]{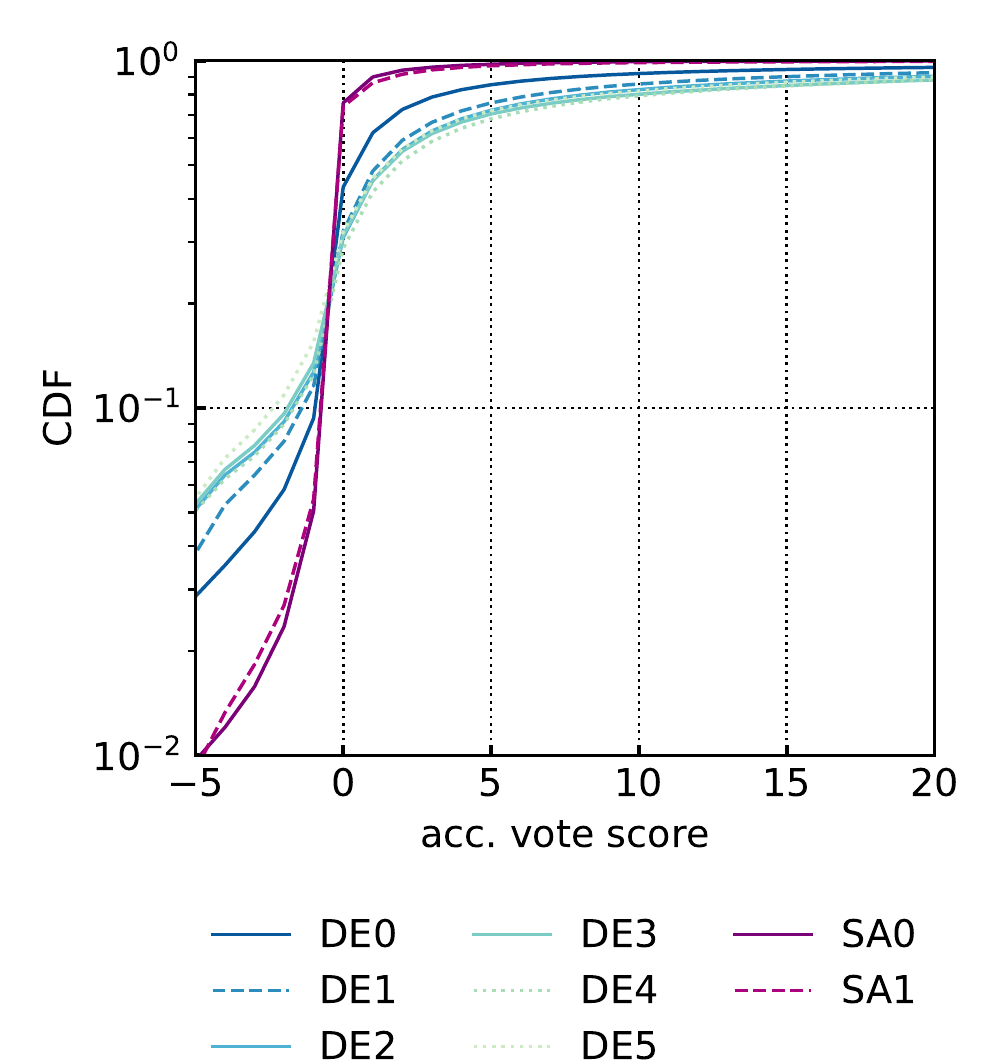}
        \caption{Accumulate vote scores CDF over time.}
        \label{fig:karma_cdf}
    \end{subfigure}
    \quad
    \begin{subfigure}[t]{.65\textwidth}
        \centering
        \includegraphics[width=\textwidth]{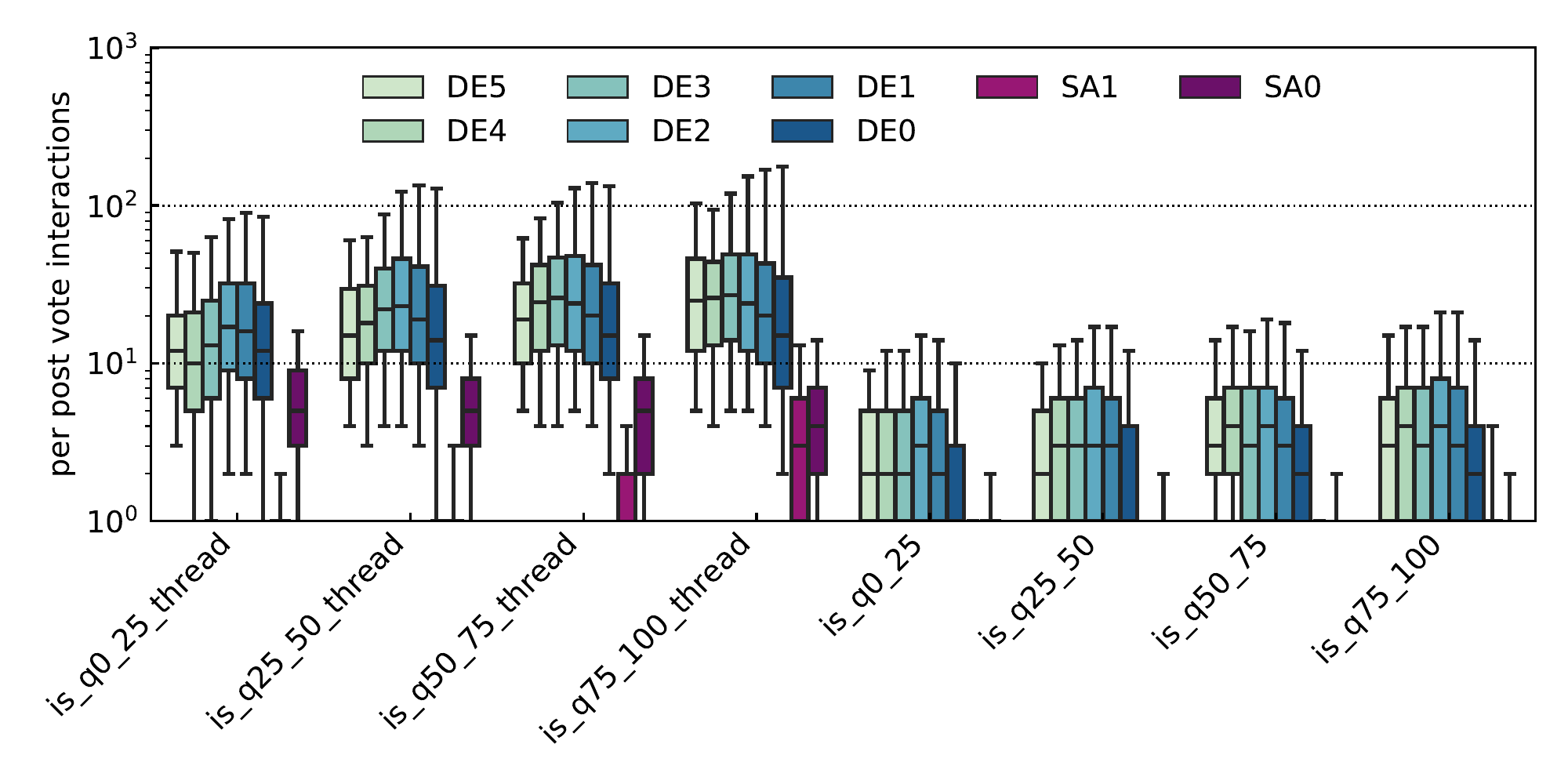}
        \caption{Vote interaction distributions by time period (DE0, ...) split by community size (q0\_25,...) and being a thread start (thread) or is a reply buried within a thread.}
        \label{fig:post_votes_boxplot}
    \end{subfigure}
    \caption{Platform accumulate vote scores and vote interactions per post. \emph{a)} Larger communities experience a stronger heavy tail in vote scores. \emph{b)} Due to their exposure, threads usually collect more votes than replies buried within them.}
    \vspace{-1.5em}
\end{figure}

\begin{figure}[t]
    \vspace*{-1em}
    \centering
    \begin{subfigure}[t]{.3\textwidth}
        \centering
        \includegraphics[width=\textwidth]{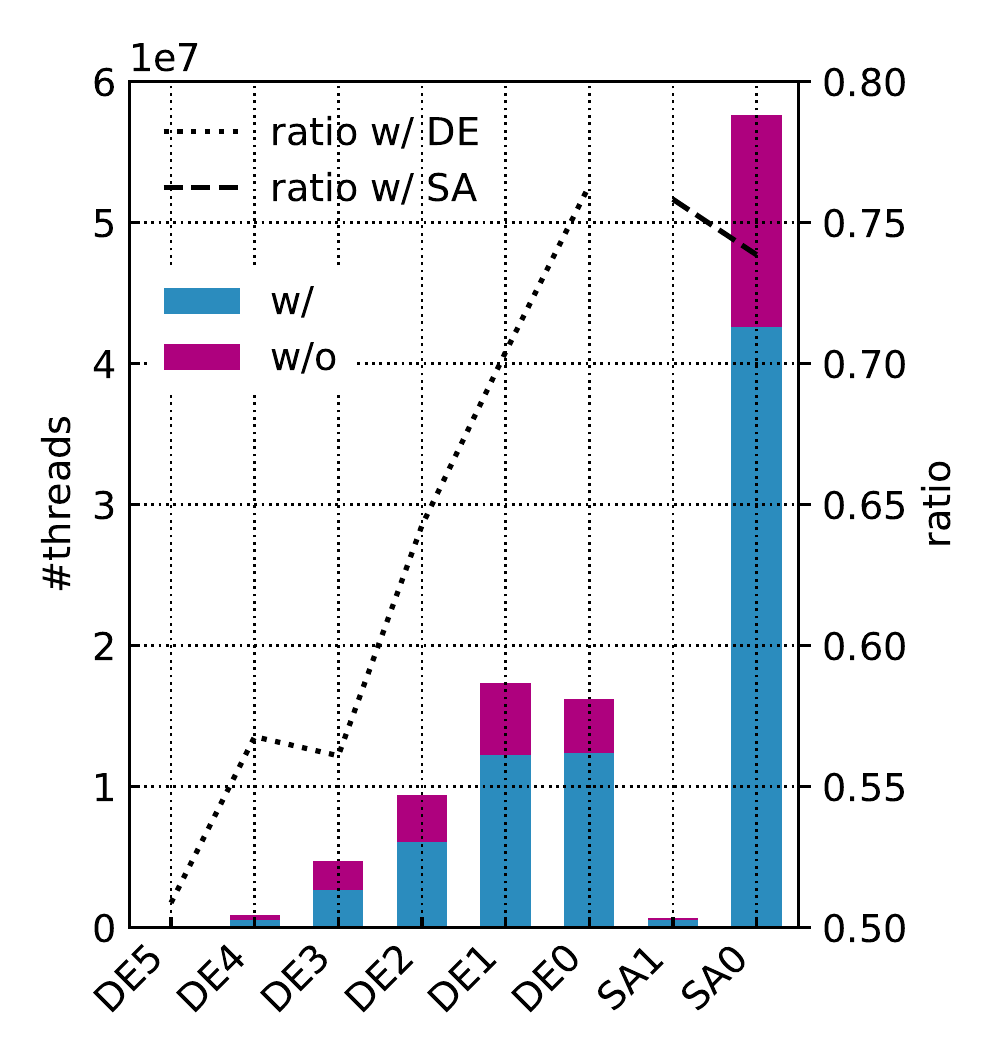}
        \caption{Thread w/o replies over time and response ratio. Most threads can attract replies.}
        \label{fig:post_has_replies_boxplot}
    \end{subfigure}
    \quad
    \begin{subfigure}[t]{.65\textwidth}
        \centering
        \includegraphics[width=\textwidth]{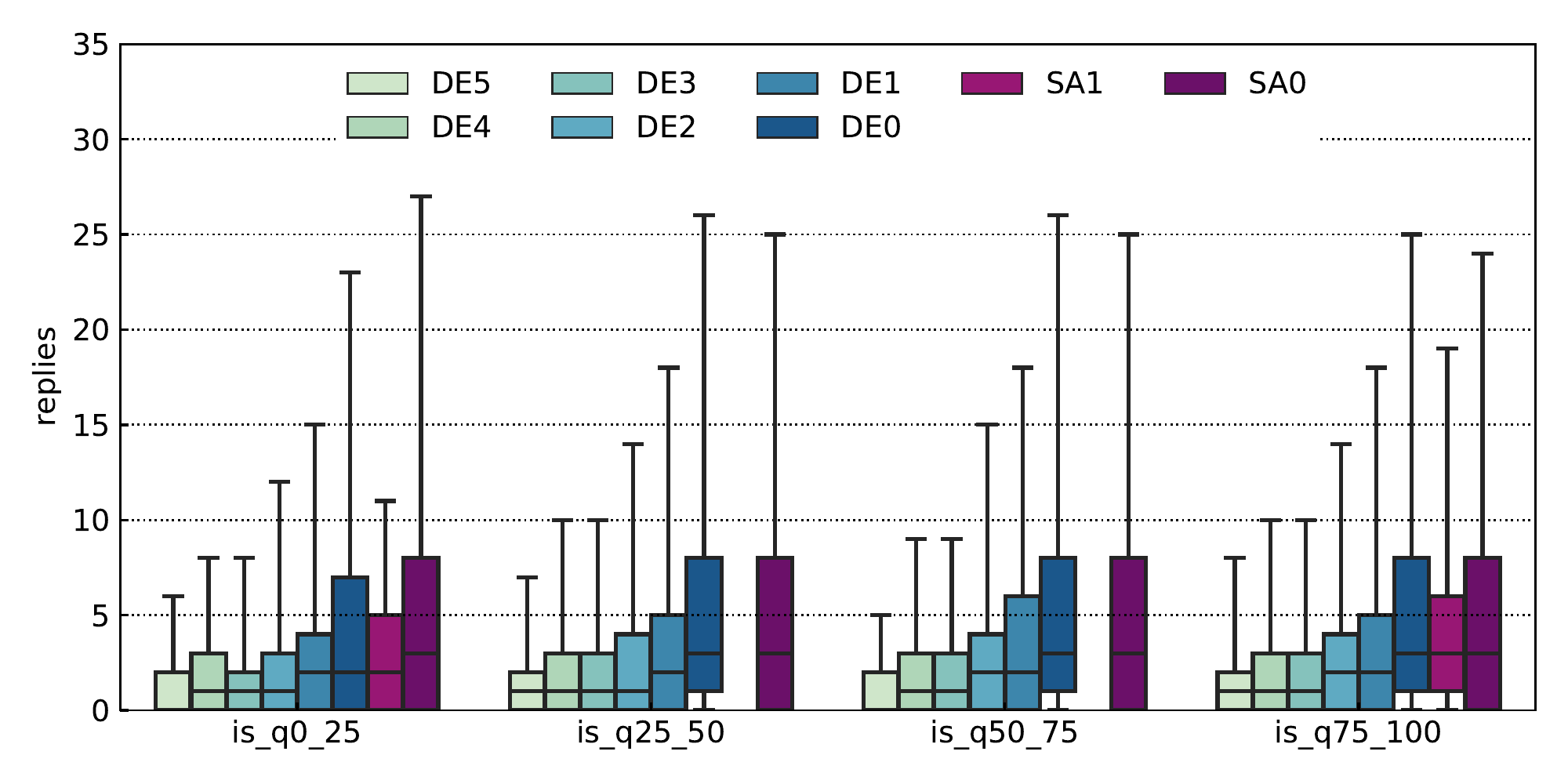}
        \caption{Thread length over time by community size. With increasing engagement (\emph{left}, Fig. \ref{fig:post_has_replies_boxplot}), communities experience longer threads on average invariant to community size.}
        \label{fig:posts_replies_boxplot}
    \end{subfigure}

    \caption{Thread engagement with responses - how many \& how much. \emph{a)} At an increasing ratio over time, within recent time periods, most threads attract replies. \emph{b)} With growing amounts of interactions over time, the average thread length increases to a similar level in DE and SA.}
    \label{fig:post_replies}
    \vspace*{-2em}
\end{figure}

\afblock{Accumulated votes overview.}
Votes on posted content have two roles in Jodel: \emph{i)} they show content appreciation to others (\eg{} enable users to sort by popularity) and \emph{ii)} enable distributed user based content moderation that removes content with negative vote scores (see \sref{sec:jodel}).
Factors influencing vote distributions can thus have structural implications on the platform.
To study differences in voting behavior, we first take a look into accumulate vote score (\#upvotes - \#downvotes) distributions.
There are two given bounds for posts gathering votes:
\emph{1)} Posts beyond a negative threshold are no more displayed on the platform.
\emph{2)} There is no conscious upper bound given by the system, yet posts are only temporarily displayed within the various app feeds (see \sref{sec:jodel}) and therefore, the time for interacting with them is inherently limited.
Given these constraints, all communities naturally enjoy a rather positive mood.
To put an emphasis on the temporal dimension, we show the CDF of vote scores to posts over time (DE0, ...) in Figure \ref{fig:karma_cdf}.
In the earlier times of DE with less activity, more posts were able to gather more votes as illustrated by the DE3..5 series resembling a broader distribution.
There is a slight decrease of accumulated vote scores throughout time, hence interactions per post, for the DE communities---also observed within the KSA.
The distributions become more and more long-tailed over time and appear scale-shifted.
Noteworthy, a split into community sizes confirms this finding: larger communities may reach far higher absolute scores, but the distributions become more skewed correlated to the observed interaction volume (not shown).

Due to SA users producing much more content, the feeds displayed in the app also get renewed completely at a very high pace.
Thus, SA posts compete harder for time to collect possible votes in comparison to their DE cousins; the feeds also promote observed long-tails.
What implications does this shift have on experienced vote distributions?

\afblock{Votes per Post.}
To better understand the voting interactions and the observed skew in accumulated vote scores, we next normalize observed figures to a per-post basis.
The box plots in Figure \ref{fig:post_votes_boxplot} show various per post vote interaction distributions across time (DE0, ...), and community size (q0\_25, ...), while further distinguishing between threads (thread) and replies.

We find the long tail of high vote scores in the rather long 95\% percentile whiskers on the log scaled y-axis. 
Invariant to time and community size, the median German user enjoys voting on threads with median levels around 10 to 30 votes gathered by each post throughout time.
Naturally given by the communication structure and app design, content buried within threads is much less appreciated; they accumulate only two to three votes in DE.
As discussed before, the SA posts stand in stark contrast at three to four votes within the main Q0 timeframe.
\Ie{} opposed to German users, the average participant within the Kingdom of Saudi Arabia cannot expect to receive any vote on her content---especially and naturally not on replies.

\takeaway{We show that the availability of more posted content in Saudi Arabia decreases the available votes per post, which can influence community moderation techniques that depend on voting.}

\subsection{Spinning Faster: Response Time and Volume}
\label{sec:structural_implications__response}
While voting or liking is a vital part of a social network, it can only exist because of posted content and replies.
We thus next study geographic properties that influence the response time and volume.

In Figure~\ref{fig:post_has_replies_boxplot}, we show the amounts of posts with and without replies (bars) and the ratio (lines) across time for both countries.
The German communities increase their response cover over time, while it instantly is equally high for the SA communities at about 90\%.
\Ie{} 9 out of 10 users in both countries can expect getting at least a single reply on a thread. 

\afblock{Response volume.}
As most users receive a reply, does the total achieved thread length correlate with community size, and how does this interplay with the distribution shift in content creation? 
We answer this question with distributions given in Fig. \ref{fig:posts_replies_boxplot} as a box plot, which depicts the thread length gathered per post across time (DE0, ...) and community sizes (is\_q0\_25, ...).

First, the 95\% percentile whiskers indicate a long-tailed distribution in the length of threads, which we confirm (now shown).
Second, the amount of replies is invariant to community size as the distributions are very similar; however, there still exists a huge spread from the 75\% to the 95\% percentile (whiskers) due to the long-tailed distribution.
Second, we observe an increasing trend over time.
This increasing engagement is also apparent when looking deeper into the interactions (cf. Section \ref{sec:partition_communities} and alike split by community size---not shown).

\takeaway{Most posts in both countries get a reply; even at larger volumes for SA, the thread lengths are similar to DE.}

\afblock{Conversations.}
Having established an understanding of the amount of replies most users experience, we get into more structural detail.
We define \emph{conservationness} as the ratio between replies per replier as a proxy for conversations---where lower ratios naturally depict a heterogeneous set of repliers, while higher ratios indicate fewer participants forming a back and forth conversation.

\begin{figure}[t]
    \vspace*{-1em}
    \centering
    \begin{subfigure}[t]{.315\textwidth}
        \centering
        \includegraphics[width=\textwidth]{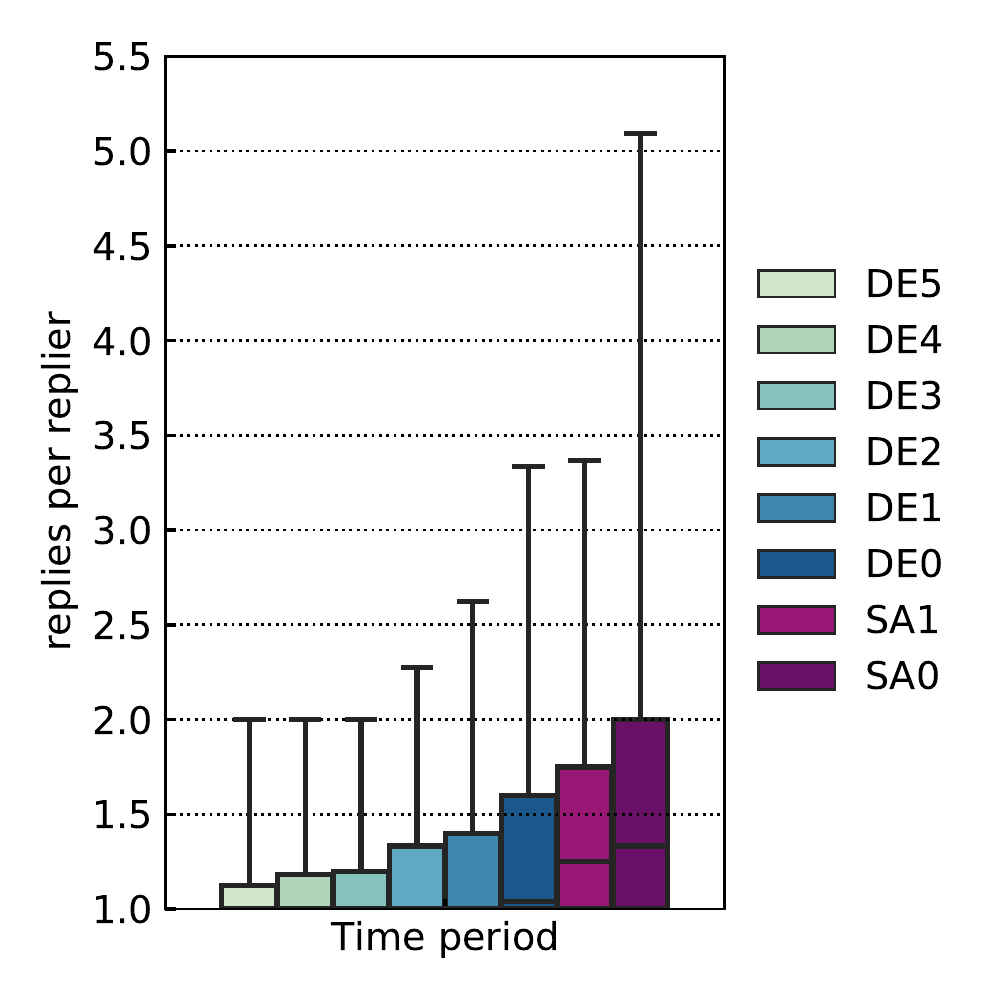}
        \caption{Conversationness - Replies per thread participant are long-tailed and substantially larger within SA.}
        \label{fig:post_repliers_replies_boxplot}
    \end{subfigure}
    \quad
    \begin{subfigure}[t]{.64\textwidth}
        \centering
        \includegraphics[width=\textwidth]{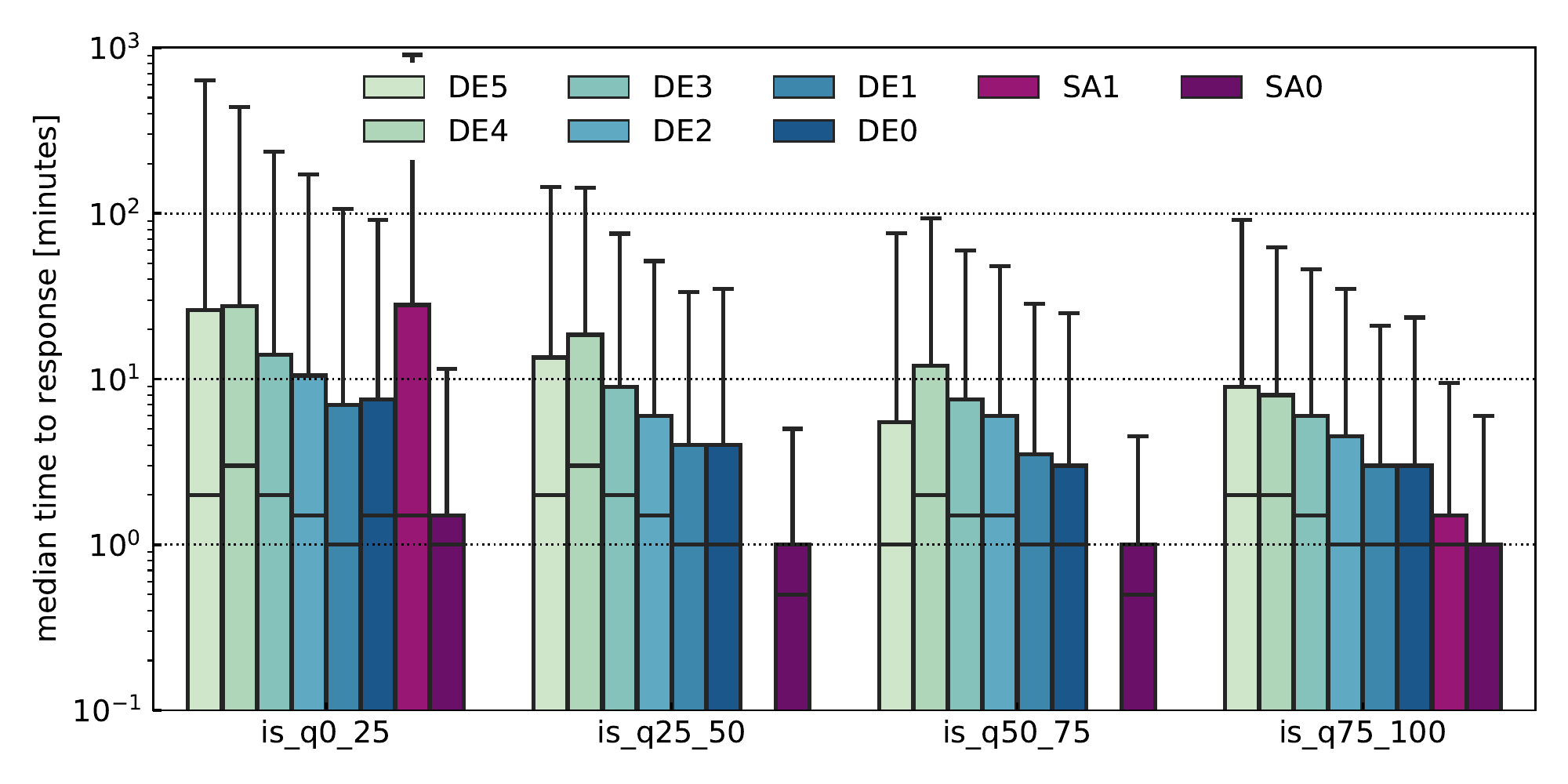}
        \caption{Thread engagement speed. The median timeframe between consecutive answers within a thread decreases over time in the meanwhile growing DE environment. Likewise, larger communities experience faster responses---in SA widely immediately.}
        \label{fig:time_to_response}
    \end{subfigure}

    \caption{Thread engagement with responses - who \& when. \emph{a)} The observed overall activity increase in DE results in longer conversations over time, the SA users talk substantially longer. \emph{b)} The average time until receiving a response reduces to only few minutes in both countries, SA still takes the lead.}
    \label{fig:replies_time}
    \vspace*{-2em}
\end{figure}

We present the distributions of this ratio over time for both countries in Figure \ref{fig:post_repliers_replies_boxplot} as box plots; the 95\% whisker indicates long-tailed distributions, which we can confirm (not shown).
Over time and with increasing network activity, all German communities increase up to about 1.6 replies per thread participant within the 75\% quantile in DE0.
This indicates a shift from rather random single comments becoming less popular in favor of interacting with each other.
Given the high preference on creating content and vividly replying, this trend is particularly apparent and reinforced in the SA communities at rates of up to two replies for the 75\% quantile of SA0.

\takeaway{Employing a conservationness metric, we identify SA users to be more conversational compared to DE.}

\afblock{Response time.}
We have seen that most threads receive at least some replies.
While the counts in responses may matter quantitatively, we also want to shed light on the time-dependent dynamics of the reply interactions.
Fig. \ref{fig:time_to_response} shows a box plot of the distributions of the time between consecutive responses within a thread split by time (DE0, ...) and community size (q0\_25, ...); note the log y-scale.
Unfortunately, our dataset does not allow for this evaluation on vote interactions (see \sref{sec:Dataset_Description_and_Statistics}).

From this evaluation, we gain two major insights: \emph{1)} throughout time with increasing activity and engagement, the German communities establish shorter response times down to only minutes.
Having reached a sustainable community size, the response times no longer drop.
\emph{2)} The SA communities instantly drop response times substantially below the German counterpart to only a single minute within most threads.
Note: High response times within small communities in SA1 are due to small amounts of data; Missing series indicate no present data.

\takeaway{In comparison to DE, the SA communities are more vividly responsive as the average response times are considerably lower.}

\section{Related Work}
\label{sec:RW}

The research community established a rich field of understanding human interaction within social media, yet not studying geographic differences in social media usage.
Empirical studies on social media focused on the birth and growth~\cite{schioberg2012tracing,mislove2008growth,reyaee2015growth}, social media usage in specific regions such as the Arab Gulf states~\cite{reyaee2015growth} specifically focused to global usage~\cite{leskovec2008}, information propagration~\cite{cha2009measurement}, specific platforms such as Facebook~\cite{lewis2008tastes,nazir2008unveiling}, YouTube~\cite{brodersen2012youtube}, SnapChat~\cite{vaterlaus2016snapchat}, or Twitter~\cite{kouloumpis2011twitter}.
Such research tries to understand and identify social structures and influence~\cite{kairam2012talking,tang2009social}.
Mathematical modelling~\cite{van2011lognormal} and graph methods are common techniques to analyze social ties~\cite{kumar2010structure,magno2012new}.
Platforms may also have rather adverse effects like cyberbullying~\cite{whittaker2015cyberbullying,kayes2015social,hosseinmardi2014towards}, or may raise privacy concerns~\cite{stutzman2013silent}.

A recent body of research aims at understanding anonymous social networks.
The desire for anonymity can result in throwaway accounts~\cite{leavitt2015throwaway} and can also manifest in anonymous self disclosures~\cite{birnholtz2015weird}.
Anonymous content platforms have been detailed \wrt{} content~\cite{papasavva2020raiders} and user behavior~\cite{bernstein20114chan,JodelChurn,correa2015many}.
Other empirical work focuses specifically on location based anonymous platforms, \eg{} Whisper~\cite{wang2014whispers} or Yik~Yak~\cite{mckenzie2015oxen,saveski2016tracking}, or analyzes its local content~\cite{black2016anonymous}.

We complement these works providing a new unique view on the lifecycle of various Jodel communities in Germany and Saudi Arabia with a special focus on happening interactions, differences and resulting platform implications.

\section{Conclusions}
In this paper, we show that the usage behavior of users in Germany (DE) fundamentally differs from users in Saudi Arabia (KSA) in the anonymous and location based Jodel network.
This study is enabled by the feature of Jodel to form independent local communities enables us to compare in-country and between country effects and thereby to clearly identify country specific usage differences.
We empirically characterizes usage behavior based on ground truth user interaction data provided by the operator.
While we can rule out marketing effects by the operator, our findings motivate future work that study root causes.
We find that, independent of time and community size, KSA users prefer content creation (posting \& responding), while German users tend to interact slightly more passively (voting).
Other than this shift towards content, due to the users in both regions else behaving identical on a per-user measure; we find rather identical community engagement.
However, due to much more content being available within SA per user, posts compete harder in gathering votes than the German counterparts, which can have implications for vote-based content moderation schemes.
Further, the average number of replies also does not increase in comparison; Still, reply times are much smaller due to higher activity.
The average Saudi user tends towards having longer conversations.
Overall, we identify time- and geographic-invariant differences between DE and SA user engagement as the latter substantially focus on creating content, giving a slight lead in voting to the Germans.
This provides a new interaction-based perspective on geographic difference of social media usage that have not yet been studied.
\bibliographystyle{splncs04}
\bibliography{references}

\end{document}